\documentclass[aps,showpacs,prb,floatfix,twocolumn]{revtex4}
\usepackage{amsmath,amssymb,graphicx,bm,epsfig,colordvi}

\input{tcilatex}

\begin{document}

\title{Interpolative Approach for Solving the Anderson Impurity Model}
\author{S. Y. Savrasov$^{a}$, V. Oudovenko$^{b,e}$, K. Haule$^{c,e}$, D.
Villani$^{d,e}$, G. Kotliar$^{e}$}
\affiliation{$^{a}$Department of Physics, New Jersey Institute of Technology, Newark, NJ
07102, USA}
\affiliation{$^{b}$Laboratory for Theoretical Physics, Joint Institute for Nuclear
Research, 141980 Dubna, Russia}
\affiliation{$^{c}$Jo\v zef Stefan Institute, SI-1000 Ljubljana, Slovenia}
\affiliation{$^{d}$JPMorgan Chase Bank, 270 Park Avenue, New York, NY 10017}
\affiliation{$^{e}$Department of Physics and Center for Material Theory, Rutgers
University, Piscataway, NJ 08854, USA}
\date{\today}

\begin{abstract}
A rational representation for the self--energy is explored to interpolate
the solution of the Anderson impurity model in general orbitally degenerate
case. Several constrains such as the Friedel's sum rule, positions of the
Hubbard bands as well as the value of quasiparticle residue are used to
establish the equations for the coefficients of the interpolation. We employ
two fast techniques, the slave--boson mean--field and the Hubbard I
approximations to determine the functional dependence of the coefficients on
doping, degeneracy and the strength of the interaction. The obtained
spectral functions and self--energies are in good agreement with the results
of numerically exact quantum Monte Carlo method.
\end{abstract}

\pacs{71.10.-w, 71.27.+a, 71.30.+h}
\maketitle

\section{Introduction}

There has been recent progress in understanding physics of strongly
correlated electronic systems and their electronic structure near a
localization--delocalization transition through the development of dynamical
mean--field theory (DMFT) \cite{DMFT}. Merging this computationally
tractable many--body technique with realistic local--density--approximation
(LDA)\ \cite{LDA} based electronic structure calculations of strongly
correlated solids is promising due to its simplicity and correctness in both
band and atomic limits. At present, much effort is being made in this
direction including the developments of a LDA+DMFT method \cite%
{AnisimovKotliar}, LDA++ approach \cite{LDA++}, combined GW and DMFT theory 
\cite{Georges}, spectral density functional theory \cite{SDFT} as well as
applications to various systems such as La$_{1-x}$Sr$_{x}$TiO$_{3}$ \cite%
{LaTiO3}, V$_{2}$O$_{3}$ \cite{V2O3}, Fe and Ni \cite{Licht}, Ce \cite{Ce},
Pu \cite{Nature,Science}, transition metal oxides\cite{NiOPRL}, and many
others. For a review, see Ref. \onlinecite{reviews}.

Such \textit{ab initio} DMFT based self--consistent electronic structure
algorithms should be able to explore all space of parameters where neither
dopings nor even degeneracy itself is kept fixed as different states may
appear close to the Fermi level during iterations towards self--consistency.
This is crucial if one would like to calculate properties of realistic solid
state system where bandwidth and the strength of the interaction is not
known at the beginning. It is very different from the ideology of model
Hamiltonians where the input set of parameters defines the regime of
correlations, and the corresponding many--body techniques may be applied
afterwards. Realistic DMFT\ simulations of material properties require fast
scans of the entire parameter space to determine the interaction for a given
doping, degeneracy and bandwidth via the solution of the general
multiorbital Anderson impurity model (AIM) \cite{AIM} . Unfortunately,
present approaches based on either non--crossing approximation (NCA)\ or
iterative perturbation theory (IPT) are unable to provide the solution to
that problem due to a limited number of regimes where these methods can be
applied \cite{DMFT}. The quantum Monte Carlo (QMC) technique \cite%
{DMFT,Jarrell} is very accurate and can cope with multiorbital situation but
not with multiplet interactions. Also its applicability so far has been
limited either to a small number of orbitals or to unphysically large
temperatures due to its computational cost. Recently some progress has been
achieved using impurity solvers that improve upon the NCA approximation \cite%
{rotors,jeschke,Haule:2001}, but it has not been possible to retrieve Fermi
liquid behavior at very low temperatures with these methods in the orbitally
degenerate case.

As universal impurity solvers have not yet being designed in the past we
need to explore other possibilities, and this paper proposes interpolative
approach for the self--energy in general multiorbital situation. We stress
that this is not an attempt to develop an alternative method for solving the
impurity problem, but follow--up of the ideology of LDA theory where
approximations were designed by analytical fits \cite{Vosko} to the quantum
Monte Carlo simulations for homogeneous electron gas \cite{Ceperley}.
Numerically very expensive QMC calculations for the impurity model display
smooth self--energies at imaginary frequencies for a wide range of
interactions and dopings, and it is therefore tempting to design such an
interpolation. We also keep in mind that for many applications a high
precision in reproducing the self--energies may not be required. One of such
applications is, for example, the calculation of the total energy \cite%
{Ce,Nature,Science,NiOPRL} which, as well known from LDA based experience,
may not be so sensitive to the details of the one--electron spectra. As a
result, we expect that even crude evaluations of the self--energy shapes on
imaginary axis may be sufficient for solving many realistic total energy
problems, some of which have appeared already \cite{Nature,Science,NiOPRL}.
Another point is a computational efficiency and numerical stability.
Bringing full self--consistent loops with respect to charge densities \cite%
{Nature} and other spectral functions require many iterations towards the
convergency which may not need too accurate frequency resolutions at every
step. However, the procedure which solves the impurity model should smoothly
connect various regions of the parameter space. This is a crucial point if
one would like to have a numerically stable algorithm and our new
interpolational approach ideally solves this problem.

In the calculations of properties such as the low--energy spectroscopy and
especially transport more delicate distribution of spectral weight is taken
place at low energies, and the imaginary part of the analytically continued
self--energy needs to be computed with a greater precision. Here we expect
that our obtained spectral functions should be used with care.{\ Also, in a
few well distinct regimes, such, e.g., as very near the Mott transition, the
behavior maybe much more complicated and more difficult to interpolate. For
the cases }mentioned above{\ extensions of the interpolative methods should
be implemented and its beyond the scope of the present work. }

We can achieve a fast interpolative algorithm for the self--energy by
utilizing a rational representation. The coefficients in this interpolation
can be found by forcing the self--energy to obey several limits and
constrains. For example, if infinite frequency (Hartree--Fock) limit,
positions of the Hubbard bands, low--frequency mass renormalization $z$,
mean number of particles $\bar{n}$ as well as the value of the self--energy
at zero frequency $\Sigma (0)$ are known from independent calculation, the
set of interpolating coefficients is well defined. In this work, we explore
the slave--boson mean--field (SBMF) approach \cite{Gutz,Ruck,
Fleszar,Hasegawa} and the Hubbard I approximation \cite{Hubbard1} to
determine the functional dependence of these coefficients upon doping,
degeneracy and the strength of the interaction $U$. We verify all trends
produced by this interpolative procedure in the regimes of weak,
intermediate and strong interactions and at various dopings conditions.
These trends are compared with known analytical limits as well as against
calculations using the quantum Monte Carlo method. Also, compared with QMC
are self--energies and spectral functions on both imaginary and real axes
for selective values of dopings. They indicate that the SBMF approach can
predict such parameters of interpolation as $\bar{n},\Sigma (0)$and $z$ with
a good accuracy while the Hubbard I method fails in a number of regimes.
However, the functional form of the atomic Green function which appears
within Hubbard I can be used to determine positions of atomic satellites
which helps to impose additional constraints on our procedure.

Giving an extraordinary computational speed of this approach we generally
find a very satisfactory accuracy in comparisons with the numerically more
accurate QMC calculations. If an increased accuracy is desired our method
can be naturally extended by imposing more constraints and by implementing
more refined impurity solvers other than the ones explored in this work.

The paper is organized as follows. In Section II we discuss rational
interpolation for the self--energy and list the constraints. In Section III
we discuss methods for solving Anderson impurity model such as the
slave--boson mean field and the Hubbard I approximations which can be used
to find these constraints. A brief survey of the QMC method used to
benchmark our algorithm is also given. We present numerical comparisons of
SBMF and Hubbard I techniques against the QMC simulations for such
quantities as quasiparticle residue and multiple occupancies. In Section IV
we report the results of the interpolative method and compare the obtained
spectral functions with the QMC. In Section V we discuss possible
improvements of the algorithm. Section VI is the conclusion.

\section{Interpolative Approach}

To be specific, we concentrate on the Anderson impurity Hamiltonian 
\begin{eqnarray}
H &=&\epsilon _{f}\sum_{\alpha =1}^{N}f_{\alpha }^{+}f_{\alpha }+\frac{1}{2}%
U\sum_{\alpha \neq \beta }^{N}n_{\alpha }^{f}n_{\beta }^{f}+\sum_{\mathbf{k}%
\alpha }E_{\mathbf{k}\alpha }c_{\mathbf{k}\alpha }^{+}c_{\mathbf{k}\alpha } 
\notag \\
&+&\sum_{\mathbf{k}\alpha }[V_{\alpha }^{\ast }(\mathbf{k})f_{\alpha }^{+}c_{%
\mathbf{k}\alpha }+V_{\alpha }(\mathbf{k})c_{\mathbf{k}\alpha }^{+}f_{\alpha
}],  \label{HAM}
\end{eqnarray}%
describing the interaction of the impurity level $\epsilon _{f}$ with bands
of conduction electrons $E_{\mathbf{k}\alpha }$ via hybridization $V_{\alpha
}(\mathbf{k})$. $U$ is the Coulomb repulsion between different orbitals in
the $f$--band. Inspired by the success of the iterative perturbation theory 
\cite{DMFT}, in order to solve the Anderson impurity model in general
multiorbital case, we use a rational interpolative formula for the
self--energy. This can be encoded into a form%
\begin{equation}
\Sigma (\omega )={}\frac{\sum_{m=0}^{M}a_{m}\omega ^{m}}{\sum_{m=0}^{M}b_{m}%
\omega ^{m}}=\Sigma (\infty )\frac{\prod\limits_{m=1}^{M}[\omega
-Z_{m}^{(\Sigma )}]}{\prod\limits_{m=1}^{M}[\omega -P_{m}^{(\Sigma )}]},
\label{RAT}
\end{equation}%
The coefficients $a_{m}$, $b_{m}$, or, alternatively, the poles $%
P_{m}^{(\Sigma )}$, zeroes $Z_{m}^{(\Sigma )}$ and $\Sigma (\infty )$ in
this equation are to be determined. The form (\ref{RAT}) can be also viewed
as a continuous fraction expansion but continuous fraction representation
will not be necessary for the description of the method.

Our basic assumption is that only a well distinct set of poles in the
rational representation (\ref{RAT}) is necessary to reproduce an overall
frequency dependence of the self--energy. Extensive experience gained from
solving Hubbard and periodic Anderson model within DMFT at various ratios of
the on--site Coulomb interaction $U$ to the bandwidth $W$ shows the
appearance of lower and upper Hubbard bands as well as renormalized
quasiparticle peak in the spectrum of one--electron excitations \cite{DMFT}.

It is clear that the Hubbard bands are damped atomic excitations and to the
lowest order approximation appear as the positions of the poles of the
atomic Green function. In the $SU(N)$ symmetry case which is described by
the Hamiltonian (\ref{HAM}), these energies are numerated by the number of
electrons occupying impurity level, i.e. $E_{n}=\epsilon _{f}n+\frac{1}{2}%
Un(n-1),$ and the atomic Green function takes a simple functional form 
\begin{equation}
G_{at}(\omega )=\sum_{n=0}^{N-1}\frac{C_{n}^{N-1}(X_{n}+X_{n+1})}{\omega
+\mu -E_{n+1}+E_{n}} ,  \label{LEH}
\end{equation}%
where $X_{n}$ are the probabilities to find an atom in configuration with $n$
electrons while combinatorial factor $C_{n}^{N-1}=\frac{(N-1)!}{n!(N-n-1)!}$
arrives due to equivalence of all states with $n$ electrons in $SU(N)$.

We can represent the atomic Green function (\ref{LEH}) using the rational
representation (\ref{RAT}), i.e. 
\begin{equation}
G_{at}(\omega )=\frac{\prod\limits_{n=1}^{N-1}[\omega -Z_{n}^{(G)}]}{%
\prod\limits_{n=1}^{N}[\omega -P_{n}^{(G)}]},  \label{GZP}
\end{equation}%
where $P_{n}^{(G)}$ are all $N$ atomic poles, while $Z_{n}^{(G)}$ denote $%
N-1 $ zeroes with $N$ being the total degeneracy. The centers of the Hubbard
bands are thus located at the atomic excitations $P_{n}^{(G)}=E_{n}-E_{n-1}-%
\mu =\epsilon _{f}-\mu -U(n-1)$. Using standard expression for the atomic
Green function $G_{at}(\omega )=1/[\omega +\mu -\epsilon _{f}-\Sigma
_{at}(\omega )]$, we arrive to a desired representation for the atomic
self--energy

\begin{equation}
\Sigma ^{(at)}(\omega )=\omega +\mu -\epsilon _{f}-\frac{\prod%
\limits_{n=1}^{N}[\omega -P_{n}^{(G)}]}{\prod\limits_{n=1}^{N-1}[\omega
-Z_{n}^{(G)}]} .  \label{SAT}
\end{equation}

Using this functional form for finite $\Delta (\omega )$ modifies the
positions of poles and zeroes via recalculating probabilities $X_{n}$ which
is equivalent to the famous Hubbard I approximation (discussed in more
detail in the next Section).

We now concentrate on the description of the quasiparticle peak which is
present in metallic state of the system. For this an extra pole and zero
have to be added in Eq (\ref{SAT}). To see this, let us consider the Hubbard
model for the $SU(N)$ case where the local Green function can be written by
the following Hilbert transform $G_{f}(\omega )=H[\omega +\mu -\epsilon
_{f}-\Sigma (\omega )].$ If self--energy lifetime effects are ignored, the
local spectral function becomes $N_{f}(\omega )=D[\omega +\mu -\epsilon
_{f}-\Sigma (\omega )]$ where $D$ is the non--interacting density of states.
The peaks of the spectral functions thus appear as zeroes in Eq. (\ref{SAT})
and in order to add the quasiparticle peak, one needs to add one extra zero
(denoted hereafter as $X)$ to the numerator in Eq. (\ref{SAT}). To make the
self--energy finite at $\omega \rightarrow \infty $ one has to also add one
more pole (denoted hereafter as $P_{1}^{(\Sigma )}$) which should appear in
the denominator or Eq. (\ref{SAT}). Furthermore, frequently the Hartree Fock
value for the self--energy can be computed separately and it is desirable to
have a parameter in the functional form (\ref{SAT}) which will allow us to
fix $\Sigma (\infty )$. An obvious candidate to be changed is that
self--energy pole in (\ref{SAT}) which is closest to $\omega $ equal zero.
Let us denote this parameter as $P_{2}^{(\Sigma )}$ and rewrite the
denominator of (\ref{SAT}) as $(\omega -P_{1}^{(\Sigma )})(\omega
-P_{2}^{(\Sigma )})\prod\limits_{n=1}^{N-2}[\omega -Z_{n}^{(G)}]$ where the
product is now extended over all zeroes of the atomic Green functions except
the one closest to zero and two extra poles $P_{1}^{(\Sigma )}$ and $%
P_{2}^{(\Sigma )}$ can control the width of the quasiparticle peak and $%
\Sigma (\infty ).$ Thus, we arrive to the functional form for the
self--energy

\begin{widetext}

\begin{equation}
\Sigma (\omega )=\omega +\mu -\epsilon _{f}-\frac{(\omega
-X)\prod\limits_{n=1}^{N}[\omega -P_{n}^{(G)}]}{(\omega
-P_{1}^{(\Sigma )})(\omega -P_{2}^{(\Sigma
)})\prod\limits_{n=1}^{N-2}[\omega -Z_{n}^{(G)}]}. \label{SZP}
\end{equation}

\end{widetext}

We are now ready to list all constrains of our interpolative scheme. To fix
the unknown coefficients $X,$ $P_{1}^{(\Sigma )},$ $P_{2}^{(\Sigma )},$ $%
P_{n}^{(G)},Z_{n}^{(G)}$ in Eq. (\ref{SZP}) and to write down the linear set
of equations for the coefficients $a_{m}$, $b_{m}$ in Eq. (\ref{RAT}). we
use the following set of conditions.

\textit{a)} \textit{Hartree Fock value} $\Sigma (\infty ).$ In the limit $%
\omega \rightarrow \infty $ the self--energy takes its Hartree--Fock form 
\begin{equation}
{\Sigma }(\infty )=U(N\ -1)\langle n\rangle .  \label{INF}
\end{equation}%
Mean level occupancy\textit{\ }$\langle n\rangle \ $is defined as a sum over
all Matsubara frequencies for the Green's function, i.e.%
\begin{equation}
\langle n\rangle =T\sum_{i\omega }G_{f}(i\omega )e^{i\omega 0^{+}},
\label{DEN}
\end{equation}%
where 
\begin{equation}
G_{f}(\omega )=\frac{1}{\omega +\mu -\epsilon _{f}-\Delta (\omega )-\Sigma
(\omega )} ,  \label{IMP}
\end{equation}%
defines the impurity Green function and $\Delta (\omega )$ is the
hybridization function.

\textit{b)} \textit{Zero--frequency value }$\Sigma (0).$ The so called
Friedel sum rule establishes the relation between the total density and the
real part of the self--energy at zero frequency%
\begin{eqnarray}
\langle n\rangle &=&\frac{1}{2}+\frac{1}{\pi }\mathrm{arctg}\left( \frac{%
\epsilon _{f}+\Re \Sigma (i0^{+})+\Re \Delta (i0^{+})}{\Im \Delta (i0^{+})}%
\right)  \notag \\
&+&\int\limits_{-i\infty }^{+i\infty }\frac{dz}{2\pi i}G_{f}(z)\frac{%
\partial \Delta (z)}{\partial z}e^{z0^{+}}.  \label{FRD}
\end{eqnarray}

\textit{c) Quasiparticle mass renormalization value }$\partial \Re \Sigma
/\partial \omega |_{\omega =0}.$ The slope of the self--energy at zero
frequency controls the quasiparticle residue, $z$ using the following
relationship 
\begin{equation}
\frac{{\partial \Re \Sigma }}{{\partial }\omega }\mid _{\omega =0}=1-z^{-1}.
\label{MAS}
\end{equation}%
Formally, constraints (\textit{b}) and (\textit{c}) hold for zero
temperature only but we expect no significant deviations in many regions of
parameters as long as we stay at low enough temperatures. The behavior may
be more elaborated in the vicinity of Mott transition \cite{Bulla}.

\textit{d)} \textit{Positions of Hubbard bands.}

As we discussed, in order that the self--energy obeys the atomic limit and
places the centers of the Hubbard bands at the positions of the atomic
excitations, we demand that

\begin{equation}
P_{n}^{(G)}+\mu -\epsilon _{f}=\Sigma (P_{n}^{(G)}) .  \label{POL}
\end{equation}%
This condition fixes almost all self--energy zeroes $Z_{m}^{(\Sigma )}$ in
Eq.(\ref{RAT}) to the poles $P_{n}^{(G)}.$ However, it alone does not ensure
that the weight is correctly distributed among the Hubbard bands and that
the very distant Hubbard bands disappear. For this to occur, distant poles
of Green function have to be canceled out by nearby zeroes of the Green
function. It is clear that each pole $P_{n}^{(G)}$ far from the Fermi level
has to be accompanied by a nearby zero $Z_{n}^{(G)}$ in order the weight of
the pole be small. Thus, the self--energy has poles at the position of
Green's function zeroes which can be encoded into the constrain 
\begin{equation}
\lbrack \Sigma (Z_{n}^{(G)})]^{-1}=0 .  \label{ZER}
\end{equation}%
We want to keep this property of the self--energy for finite $\Delta (\omega
)$ and thus demand that self--energy diverges (when lifetime effects are
kept, it only reaches a local maximum) at the zeros of the functional form (%
\ref{LEH}) of $G_{at}(\omega ).$ Note that the relationship (\ref{ZER})
holds (approximately) for frequency $\omega $ larger than the renormalized
bandwidth $zW.$ Therefore the information about one $Z_{n}^{(G)} $ which
lies close to $\omega =0$ is omitted and replaced by the information about $%
\Sigma (\infty ),$ $\Sigma (0)$ and $z$ as it is done by separating $%
P_{1}^{(\Sigma )}$ and $P_{2}^{(\Sigma )}$ in the denominator of Eq. (\ref%
{SZP}).

We can now write down\ a set of linear equations for all unknown
coefficients in the expression (\ref{RAT}). There is total $2M+2$ of
parameters $a_{m}$ and $b_{m},m=0,M,$ where we can always set $b_{0}=1.$ The
conditions \textit{a)},\textit{b),c) }give 
\begin{eqnarray}
a_{0} &=&\Sigma (0) ,  \label{CO1} \\
b_{0} &=&1 ,  \label{CO2} \\
a_{1}-\Sigma (0)b_{1} &=&1-z^{-1} ,  \label{CO3} \\
a_{M}-b_{M}\Sigma (\infty ) &=&0 .  \label{CO4}
\end{eqnarray}%
According to condition \textit{d)} we can use all $N$ poles $P_{n}^{(G)}$
and $N-2$ zeroes $Z_{n}^{(G)}.$ The zero $Z_{n}^{(G)}$ closest to $\omega =0$
is dropped out. This brings additional $2N-2$ equations for the coefficients
and makes $M=N$ as the degree of the rational interpolation which are
written below

\begin{widetext}

\begin{eqnarray}
{}\sum_{m=0}^{N}a_{m}[P_{n}^{(G)}]^{m}-(P_{n}^{(G)}+\mu -\epsilon
_{f})\sum_{m=0}^{N}b_{m}[P_{n}^{(G)}]^{m} &=&0\text{ for }n=1...N
,
\label{CO5} \\
\sum_{m=0}^{N}b_{m}[Z_{n}^{(G)}]^{m} &=&0\text{ for }n=1...N-2
.\label{CO6}
\end{eqnarray}

\end{widetext}

Note that while $M$ may be rather large, the actual number of poles
contributing to the self--energy behavior is indeed very small. We can
directly see this from Eq. (\ref{SAT}) which uses all $N$ poles $P_{n}^{(G)}$
fulfilling Eq. (\ref{POL}) and uses $N-2$ zeroes $Z_{n}^{(G)}$ directly
related to $N-2$ poles $P_{m}^{(\Sigma )}.$ Clearly, when the spectral
weight of the atomic excitation becomes small, the corresponding $%
P_{n}^{(G)} $ becomes close to $Z_{n}^{(G)}$ and the cancellation occurs.
Therefore in realistic situations when only the upper and lower Hubbard
bands have significant spectral weight along with the quasiparticle peak,
the actual degree of the polynomial expansion is either two or three.
However, it is advantageous numerically and cheap computationally to keep
all poles and zeroes in Eq. (\ref{SZP}) because the formula automatically
distributes spectral weight over all existing Hubbard bands.

In the limit when $U\rightarrow 0$ the self--energy automatically translates
to the non--interacting one. The atomic poles get close to each other but,
most importantly, their spectral weight goes rapidly to zero as it gets
accumulated within the quasiparticle band.

In the Mott insulating regime, the conditions \textit{b)} and \textit{c)}
drop out while all poles $P_{n}^{(G)}$ and zeroes $Z_{n}^{(G)}$ can be used
to determine the interpolation. However, in this regime it does not matter
whether one of $Z_{n}^{(G)}$ closest to $\omega =0$ is dropped out or kept,
since we can always replace this information by information about $\Sigma
(\infty ).$ Therefore the Mott transition can be studied without changing
the constraints.

We thus see that in the insulating case the self--energy correctly
reproduces the well--known result of the Hubbard I\ method where the Green
function is computed after Eq. (\ref{IMP}) with atomic self--energy. If the
lifetime effects are computed, the parameters $P_{n}^{(G)}$ and $Z_{n}^{(G)}$
become complex and the Hubbard bands will acquire an additional bandwidth.
This effect is evident from the simulations using various perturbative or
QMC impurity solvers and can be naturally incorporated into the
interpolative formulas (\ref{RAT}) or (\ref{SZP}). However, in practical
implementation below we will omit it for illustrative purposes.

Let us now discuss the quality of interpolation from the perspective of the
high--frequency behavior for the self--energy. The latter can be viewed \cite%
{Moments} as expansion in terms of the moments $\Sigma ^{(m)}$, i.e., 
\begin{equation*}
\Sigma (\omega \rightarrow \infty )=\sum_{m=0}^{\infty }\frac{\Sigma ^{(m)}}{%
\omega ^{m}},
\end{equation*}%
.

Most important for us is to look at highest moments which are given by the
Hartree Fock value, Eq. (\ref{INF}) involving single occupancy matrix $%
\langle n\rangle $, as well as the first moment

\begin{equation}
\Sigma ^{(1)}=[(N-1)(N-2)\langle nn\rangle +(N-1)\langle n\rangle
-(N-1)^{2}\langle n\rangle ^{2}]U^{2},  \label{MS1}
\end{equation}%
containing a double occupancy matrix $\langle nn\rangle .$ We see that the
interpolation in part relies on the accuracy in computing multiple
occupancies which are the functionals of both atomic excitations and the
hybridization function. In this regard, using exact atomic Green function to
find poles $P_{n}^{(G)}$ and zeroes $Z_{n}^{(G)}$ as part of the constrained
procedure may not be as accurate since it would assume the use of\textit{\ }%
\emph{atomic} multiple occupancies which do \emph{not} carry information
about $\Delta (\omega ).$On the other hand, we can also use only a
functional form of the atomic Green function where the multiple occupancies
are computed in a more rigorous manner. In the next Section we will show how
this can be implemented using the SBMF multiple occupancies which will be
found to be in better agreement with the quantum Monte Carlo data.

Note that the moments $\Sigma ^{(m)}$ themselves can be used in establishing
the constraints for interpolation coefficients. This would involve
independent evaluations of $\langle n\rangle ,\langle nn\rangle ,\langle
nnn\rangle $, etc. as well as various integrals involving hybridization
function $\Delta (\omega )$. However, we may run into ill--defined numerical
problem since high--frequency information will be used to extract the
low--frequency behavior. Therefore, it is more advantageous numerically to
use some poles and zeros of $G^{at}(\omega )$ as given by condition \textit{%
d)} above.

We thus see that the interpolational scheme is defined completely once a
prescription for obtaining parameters such as $\Sigma (0),z,$ $\langle
n\rangle $ as well as poles $P_{n}^{(G)}$, and zeroes $Z_{n}^{(G)}$ is
given. For this purpose we will test two commonly used methods: SBMF method
due to Gutzwiller \cite{Gutz} as described by Kotliar and Ruckenstein \cite%
{Ruck} and the well--known Hubbard I approximation \cite{Hubbard1}. We
compare these results against more accurate but computationally demanding
quantum Monte Carlo calculations and establish the procedure to extract all
necessary parameters.

Note that once the constraints such as $z$ are computed from a given
approximate method, some of the quantities such as the total number of
particles, $\langle n\rangle $, and the value of the self--energy at zero
frequency, $\Sigma (0),$ can be computed fully self--consistently. They can
be compared with their non--self--consistent values. If the approximate
scheme already provides a good approximation for $\langle n\rangle $ and
satisfies the Friedel sum rule, the self--consistency can be avoided hence
accelerating the calculation. Indeed we found that inclusion of the
self--consistency improves the results only marginally except when we are in
a close vicinity to the Mott transition but here we do not expect that our
simple interpolative algorithm is very accurate.

We now give the description of the approximate methods for solving the
impurity model and then present the comparisons of our interpolative
procedure with the QMC calculations.

\section{Methods for Solving Impurity Model}

\subsection{\textit{Quantum Monte Carlo Method}}

The quantum Monte Carlo method is a powerful and manifestly not perturbative
approach in either interaction $U$ or the bandwidth $W$. In the QMC method
one introduces a Hubbard--Stratonovich field and averages over it using the
Monte Carlo sampling. This is a controlled approximation using different
expansion parameter, the size of the mesh for the imaginary time
discretization. Unfortunately it is computationally very expensive as the
number of time slices and the number of Hubbard--Stratonovich fields
increases. Also the method works best at imaginary axis while analytical
continuation is less accurate and has to be done with a great care.
Extensive description of this method can be found in Ref. \onlinecite{DMFT}.
We will use this method to benchmark our calculations using approximate
algorithms described later in this Section.

\subsection{\textit{Slave--Boson Mean Field Method}}

A fast approach to solve a general impurity problem is the slave--boson
method\ \cite{Ruck, Fleszar,Hasegawa}. At the mean field level, it gives the
results similar to the famous Gutzwiller approximation \cite{Gutz}. However,
it is improvable by performing fluctuations around the saddle point. This
approach is accurate as it has been shown recently to give the exact
critical value of $U$ in the large degeneracy limit\ at half--filling \cite%
{Florens}.

The main idea is to rewrite atomic states consisting of $n$ electrons $%
\left\vert \gamma _{1},...,\gamma _{n}\right\rangle $, $0\leq n\leq N$ with
help of a set of slave--bosons $\{\psi _{n}^{\gamma _{1,...,}\gamma _{n}}\}$%
. In the following, we assume $SU(N)$ symmetric case, i.e., equivalence
between different states $\left\vert \gamma _{1},...,\gamma
_{n}\right\rangle $ for fixed $n$. Formulae corresponding to a more general
crystal--field case are given in Appendix B. The creation operator of a
physical electron is expressed via slave particles in the standard manner\ 
\cite{Fleszar}. In order to recover the correct non--interacting limit at
the mean--field level, the Bose fields $\psi _{n\text{ }}$ can be considered
as classical values found from minimizing a Lagrangian $L\{\psi _{n}\}$
corresponding to the Hamiltonian (\ref{HAM}). Two Lagrange multipliers $%
\lambda $ and $\Lambda $ should be introduced in this way, which correspond
to the following two constrains: 
\begin{eqnarray}
&&\sum_{n=0}^{N}C_{n}^{N}\psi _{n}^{2}=1,  \label{ONE} \\
&&\sum_{n=0}^{N}nC_{n}^{N}\psi _{n}^{2}=TN\sum_{i\omega }G_{g}(i\omega
)e^{i\omega 0^{+}}=\bar{n}.  \label{TWO}
\end{eqnarray}

The numbers $\psi _{n}^{2}$ are similar to the probabilities $X_{n}$
discussed in connection to the formula for the atomic Green function (\ref%
{LEH}). We thus see the physical meaning of the first constrain which is the
sum of probabilities to find atom in any state is equal to one, and the
second constrain gives the mean number of electrons coinciding with that
found from $G_{g}(\omega )=(\omega -\lambda -b^{2}\Delta (i\omega ))^{-1}$.
A combinatorial factor $C_{n}^{N}=\frac{N!}{n!(N-n)!}$ arrives due to
assumed equivalence of all states with $n$ electrons.

Minimization of $L\{\psi _{n}\}$ with respect to $\psi _{n}$ leads us to the
following set of equations to determine the quantities $\psi _{n}$: 
\begin{equation*}
\lbrack E_{n}+\Lambda -n\lambda ]\psi _{n}+nbT\sum_{i\omega }\Delta (i\omega
)G_{g}(i\omega )[LR\psi _{n-1}+\psi _{n}bL^{2}]+
\end{equation*}%
\begin{equation}
+(N-n)bT\sum_{i\omega }\Delta (i\omega )G_{g}(i\omega )[R^{2}b\psi
_{n}+LR\psi _{n+1}]=0,  \label{GUZ}
\end{equation}%
where $b=RL\sum_{n=1}^{N}C_{n-1}^{N-1}\psi _{n}\psi _{n-1}$, determines the
mass renormalization, and the coefficients $L=(1-\sum_{n=1}^{N}C_{n-1}^{N-1}%
\psi _{n}^{2})^{-1/2}$, $R=(1-\sum_{n=0}^{N}C_{n}^{N-1}\psi _{n}^{2})^{-1/2}$
are normalization constants as in Refs. \onlinecite{Ruck, Fleszar}. $%
E_{n}=\epsilon _{f}n+Un(n-1)/2$ is the total energy of the atom with $n$
electrons in $SU(N)$ approximation.

Eq. (\ref{GUZ}), along with the constrains (\ref{ONE}), (\ref{TWO})
constitute a set of non--linear equations which have to be solved
iteratively. In practice, we consider Eq. (\ref{GUZ}) as an eigenvalue
problem with $\Lambda $ being the eigenvalue and $\psi _{n}$ being the
eigenvectors of the matrix. The physical root corresponds to the lowest
eigenvalue of $\Lambda $ which gives a set of $\psi _{n}$ determining the
mass renormalization $Z=b^{2}.$ Since the matrix to be diagonalized depends
non--linearly on $\psi _{n}$ via the parameters $L,R,$ and $b$ and also on $%
\lambda $, the solution of the whole problem assumes the self--consistency:
i) we build an initial approximation to $\psi _{n}$ (for example the
Hartree--Fock solution) and fix some $\lambda $, ii) we solve eigenvalue
problem and find new normalized $\psi _{n}$, iii) we mix new $\psi _{n}$
with the old ones using the Broyden method~\cite{Broyden} and build new $%
L,\,R,$ and $b$. Steps ii) and iii) are repeated until the self--consistency
with respect to $\psi _{n}$ is reached. During the iterations we also vary $%
\lambda $ to obey the constrains. The described procedure provides a stable
computational algorithm for solving AIM and gives us an access to the
low--frequency Green's function and the self--energy of the problem via
knowledge of the slope of $\Im \Sigma (i\omega )$ and the value $\Re \Sigma
(0)$ at zero frequency.

\begin{figure}[tbp]
\includegraphics[height=0.5\textwidth]{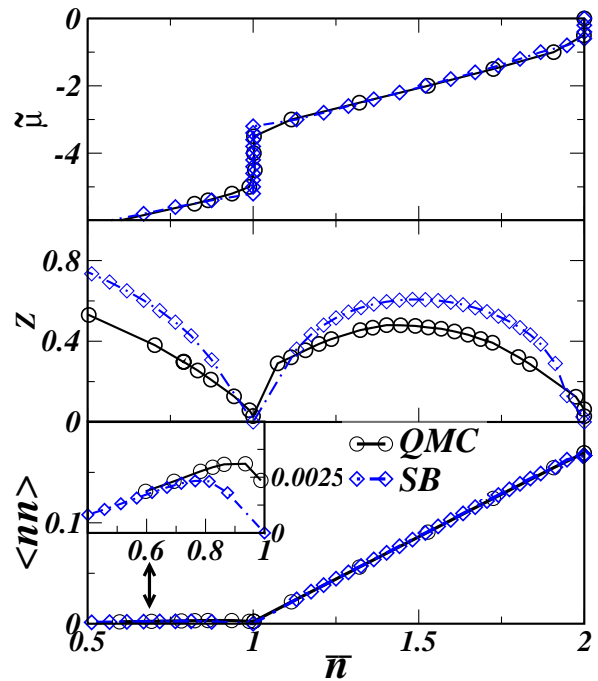}\\[-0.3cm]
\caption{Comparison between the slave boson mean field and the QMC
calculation for (a) concentration versus chemical potential $\tilde{\protect%
\mu}=\protect\mu -\protect\epsilon _{f}-(N-1)U/2$, (b) dependence of the
spectral weight $Z$ on concentration, and (c) density--density correlation
function, $\langle nn\rangle $ versus filling, $\bar{n}$, in the two--band
Hubbard model in $SU(4)$ and $U=4=2W.$}
\label{FigSBMF}
\end{figure}

The described slave--boson method gives the following expression for the
self--energy: 
\begin{equation}
\Sigma (\omega )=(1-b^{-2})\omega -\epsilon _{f}+\lambda b^{-2}.  \label{SIG}
\end{equation}%
The impurity Green function $G_{f}(\omega )$ in this limit is given by the
expression 
\begin{equation}
G_{f}(\omega )=b^{2}G_{g}(\omega ).  \label{GGF}
\end{equation}

As an illustration, we now give the solution of Eq. (\ref{GUZ}) for
non--degenerate case ($N=2)$ and at the particle--hole symmetry point with $%
\epsilon _{f}-\mu =-\frac{U}{2}(N-1)$. Consider a dynamical mean--field
theory for the Hubbard model which reduces the problem to solving the
impurity model subject to the self--consistency condition with respect to $%
\Delta (\omega )$. Starting with the semicircular density of states (DOS),
the self--consistency condition is given by Eq. (\ref{GUZ})$.$ We obtain the
following simplifications: $L=R=\sqrt{2}$, $\lambda =0$, $\psi _{0}=\psi
_{2},b=4\psi _{1}\psi _{2}$ and $G_{g}(\omega )=[\omega -(\frac{W}{4}
)^{2}b^{2}G_{g}(\omega )]^{-1}.$ The sum $T\sum_{i\omega }\Delta (i\omega
)G_{g}(i\omega )$ appeared in Eq.~(\ref{GUZ}) scales as $W\alpha /2$ with
the constant $\alpha $ being the characteristic of a particular density of
states and approximately equal to --0.2 in the semicircular DOS case. The
self--consistent solution of Eq. (\ref{GUZ}) is therefore possible and
simply gives $\psi _{2}^{2}=\frac{U}{32W\alpha }+\frac{1}{4}$. The Mott
transition occurs when no sites with double occupancies can be found, i.e.
when $\psi _{0}=\psi _{2}=0.$ A critical value of $U_{c}=8W|\alpha |$. For $%
\alpha \approx -0.2$, this gives $U_{c}\approx 1.6W$ and reproduces the
result for $U_{c2}=1.49W$ known from the QMC calculation within a few
percent accuracy. As degeneracy increases, critical $U$ is shifted towards
higher values \cite{Florens}. From numerical calculations we obtained the
following values of the critical interactions in the half--filled case $%
U_{c}\approx 3W$ for $N=6$ ($p$--level)$,$ $U_{c}\approx 4.5W$ for $N=10$ ($%
d $--level)$,$ and $U_{c}\approx 6W$ for $N=14$ ($f$--level).

Density--density correlation function $\langle nn\rangle $ for local states
with $n$ electrons is proportional to the number of pairs formed by $n$
particles $C_{2}^{n}/C_{2}^{N}$. Since the probability for $n$ electrons to
be occupied is given by: $P_{n}=\psi _{n}^{2}C_{n}^{N}$, the physical
density--density correlator can be deduced from: $\langle nn\rangle
=\sum_{n}C_{2}^{n}/C_{2}^{N}P_{n}$. Similarly, the triple occupancy can be
calculated from $\langle nnn\rangle =\sum_{n}C_{3}^{n}/C_{3}^{N}P_{n}$.

Let us now check the accuracy of this method by comparing its results with
the QMC data. We consider the two--band Hubbard model in $SU(N=4)$ orbitally
degenerate case. Hybridization $\Delta (\omega )=\sum_{\mathbf{k} }V(\mathbf{%
k})/(\omega -E_{\mathbf{k}})$ satisfies the DMFT\ self--consistency
condition of the Hubbard model on a Bethe lattice 
\begin{equation}
\Delta (\omega )=\left( \frac{W}{4}\right) ^{2}G(\omega ),  \label{SCF}
\end{equation}%
The Coulomb interaction is chosen to be $U=2W$ which is sufficiently large
to open the Mott gap at integer fillings. All calculations are done for the
temperature $T=1/32W$.

We first compare the average number of electrons vs. chemical potential
determined from the slave bosons which is plotted in Fig.~\ref{FigSBMF}(a).
This quantity is sensitive to the low--frequency part of the Green function
which should be described well by the present method. We see that it
reproduces the QMC data with a very high accuracy and only differs by 20 per
cent very near the jump of $\tilde{\mu}$ at $n=1$.

The quasiparticle residue $z$ versus filling $\bar{n}$ is plotted in Fig. %
\ref{FigSBMF}(b). The slave--boson method gives the Fermi liquid and
provides estimates for the quasiparticle residue with the overall
discrepancy of the order of 30\%. In fact, we have performed several
additional calculations for other degeneracies ($N=2$ and $6$) and for
various parameters regimes. The trend to overestimate mass renormalization
can be seen in many cases. It disappears only when $U$ approaches zero. We
need to point out, however, that i) the extractions of zero frequency
self--energy slopes from the high--temperature QMC is by itself numerically
not well grounded procedure, as information for the self--energy is known at
the Matsubara points only, which is then extrapolated to $\omega =0$, ii)
other methods for solving impurity model, such as NCA or IPT display similar
discrepancies and iii) recent findings \cite{Florens} suggest that at least
at half--filling quasiparticle residues deduced from slave bosons becomes
exact when $N\rightarrow \infty $. Most importantly for our interpolative
method is that the entire functional dependence of $z$ vs. filling,
interaction and degeneracy is correctly captured. Its overall accuracy is
acceptable as it is evident from our comparisons of the spectral functions
presented in the next Section and well within the main goal of our work to
provide fast scans of the entire parameter space necessary for simulating
real materials. This is important as, for example, for general $f$--electron
material, the QMC method is prohibitively time consuming, but we expect from
the SBMF method the results for mass renormalization not worse than 50\% for
such delicate regime as the vicinity of the Mott transition.

Fig.~\ref{FigSBMF}(c) shows the density--density correlation function $%
\langle nn\rangle $ as a function of average occupation $\bar{n}$. The
discrepancy is most pronounced for fillings $\bar{n}<1$ [see the inset of
Fig. \ref{FigSBMF}(c)] where the absolute values of $\langle nn\rangle $ are
rather small. Although our slave--boson technique captures only the
quasiparticle peak, it gives the correlation function in reasonable
agreement with the QMC for dopings not too close to the Mott transition.

\subsection{\textit{Hubbard I Approximation}}

Now we turn to the Hubbard\ I approximation\ \cite{Hubbard1} which is
closely related to the moments expansion method\ \cite{Moments}. Consider
many--body atomic states $|\Phi _{\kappa }^{(n)}\rangle $ which in $SU(N)$
are all degenerate with index $\kappa $ numerating these states for a given
number of electrons $n.$ The impurity Green function is defined as the
average%
\begin{equation}
G_{f}(\tau )=-\langle T_{\tau }f_{\alpha }(\tau )f_{\beta }^{+}(0)\rangle .
\label{GIM}
\end{equation}%
and becomes diagonal with all equal elements in $SU(N).$It is convenient to
introduce the Hubbard operators 
\begin{equation}
\hat{X}_{\kappa \kappa ^{\prime }}^{nn^{\prime }}=|\Phi _{\kappa
}^{(n)}\rangle \langle \Phi _{\kappa ^{\prime }}^{(n^{\prime })}|
\label{XOP}
\end{equation}%
and represent the one--electron creation and destruction operators as follows%
\begin{eqnarray}
f_{\alpha } &=&\sum_{n}\sum_{\kappa \kappa ^{\prime }}\langle \Phi _{\kappa
}^{(n)}|f_{\alpha }|\Phi _{\kappa ^{\prime }}^{(n+1)}\rangle \hat{X}_{\kappa
\kappa ^{\prime }}^{nn+1},  \label{FCR} \\
f_{\alpha }^{+} &=&\sum_{n}\sum_{\kappa \kappa ^{\prime }}\langle \Phi
_{\kappa }^{(n+1)}|f_{\alpha }^{+}|\Phi _{\kappa ^{\prime }}^{(n)}\rangle 
\hat{X}_{\kappa \kappa ^{\prime }}^{n+1n}.  \label{FDE}
\end{eqnarray}%
The impurity Green function (\ref{GIM}) is given by 
\begin{equation}
G_{f}(\tau )=\sum_{nm}G_{nm}(\tau ),  \label{HUB}
\end{equation}

\begin{widetext}

where the matrix $G_{nm}(\tau )$ is defined as
\begin{equation}
G_{nm}(\tau )=-\sum_{\kappa _{1}\kappa _{2}\kappa
_{3}\kappa _{4}}\langle \Phi _{\kappa _{1}}^{(n)}|f_{\alpha }|\Phi _{\kappa
_{2}}^{(n+1)}\rangle \langle T_{t}\hat{X}_{\kappa _{1}\kappa _{2}}^{nn+1}(\tau )%
\hat{X}_{\kappa _{3}\kappa _{4}}^{m+1m}(0)\rangle \langle \Phi _{\kappa
_{3}}^{(m+1)}|f_{\alpha }^{+}|\Phi _{\kappa _{4}}^{(m)}\rangle .  \label{GOP}
\end{equation}

\end{widetext}

Establishing the equations for $G_{nm}(\tau )$ can be performed using the
method of equations of motion for the $X$ operators. Performing their
decoupling due to Hubbard \cite{Hubbard1,Roth}, carrying out the Fourier
transformation and analytical continuation to the real frequency axis, and
summing over $n$ and $m$ after (\ref{HUB}) we arrive to the net result 
\begin{equation}
G_{f}^{-1}(\omega )=G_{at}^{-1}(\omega )-\Delta (\omega ),  \label{H1E}
\end{equation}%
The $G_{at}(\omega )$ can be viewed in the matrix form (\ref{HUB}) with the
following definition of a diagonal atomic Green function%
\begin{equation}
G_{nm}^{at}(\omega )=\delta _{nm}\frac{C_{n}^{N-1}(X_{n}+X_{n+1})}{\omega
+\mu -E_{n+1}+E_{n}}.  \label{ATG}
\end{equation}%
with $E_{n}=\epsilon _{f}n+Un(n-1)/2$ being the total energies of the atom
with $n$ electrons in $SU(N).$ The coefficients $X_{n}$ are the
probabilities to find atom with $n$ electrons and were already discussed in
connection to the formula (\ref{LEH}) for the atomic Green function. They
are similar to the coefficients $\psi _{n}^{2}$ introduced within the SBMF
method but now found from different set of equations. These numbers are
normalized to unity, $\sum_{n=0}^{N}C_{n}^{N}X_{n}=%
\sum_{n=0}^{N-1}C_{n}^{N-1}(X_{n}+X_{n+1})=1,$ and are expressed via
diagonal elements of $G_{nm}(i\omega )$ as follows: 
\begin{equation}
X_{n}=-T\sum_{i\omega }G_{nn}(i\omega )e^{-i\omega 0^{+}}/C_{n}^{N-1}.
\label{SCW}
\end{equation}%
Their determination in principle assumes solving a non--linear set of
equations while determining $G_{f}(\omega ).$ The mean number of electrons
can be measured as follows: $\bar{n}=\sum_{n=0}^{N}nC_{n}^{N}X_{n}$ or as
follows $\bar{n}=TN\sum_{i\omega }G_{f}(i\omega )e^{i\omega 0^{+}}.$ The
numbers $X_{n}$ can be also used to find the averages $\langle nn\rangle ,$ $%
\langle nnn\rangle $ in the way similar to what has been done in the SBMF
approach.

\begin{figure}[tbp]
\includegraphics[height=0.5\textwidth]{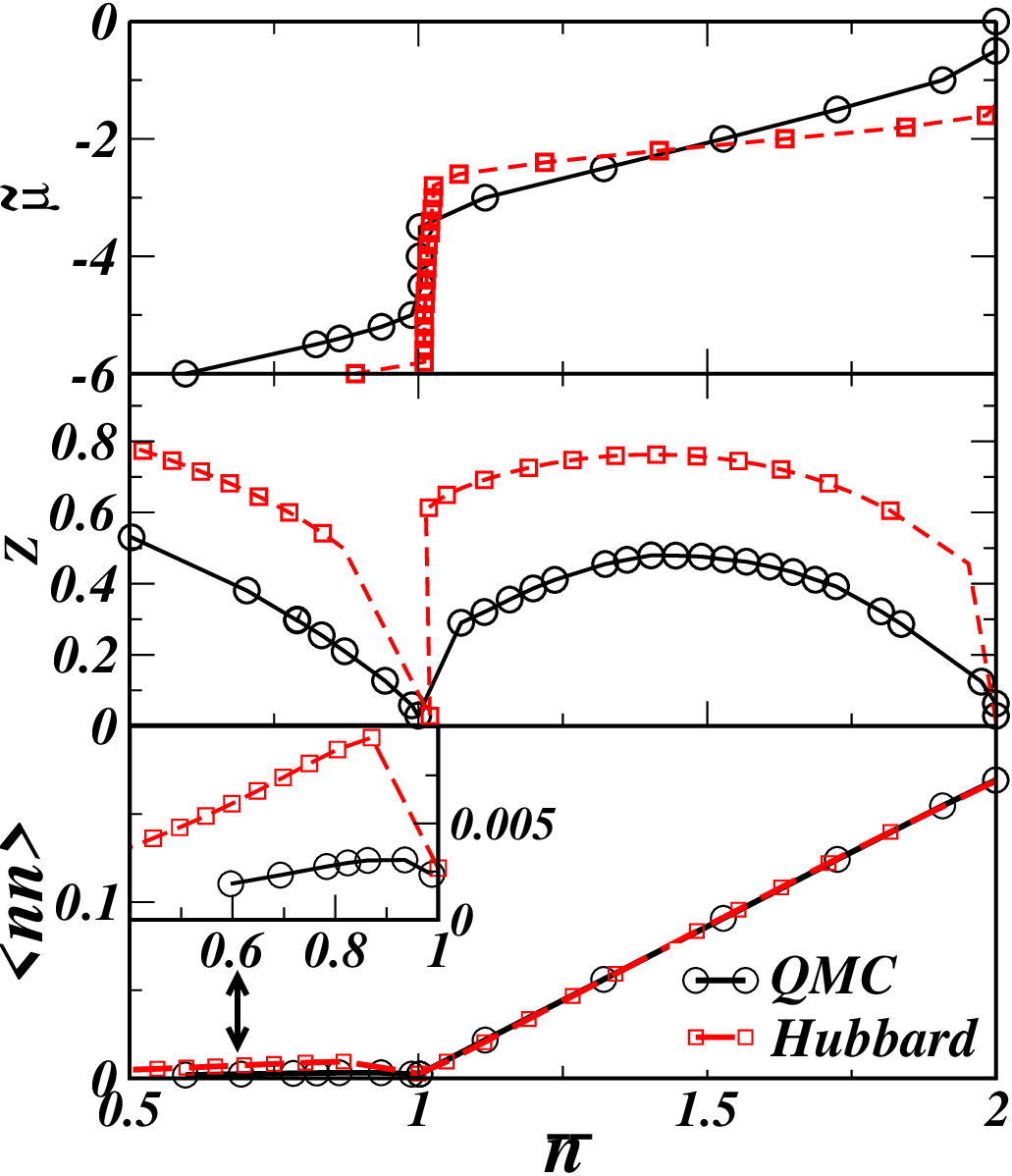}\\[-0.3cm]
\caption{Comparison between the Hubbard I and the QMC calculation for (a)
concentration versus chemical potential $\tilde{\protect\mu}=\protect\mu -%
\protect\epsilon _{f}-(N-1)U/2$, (b) dependence of the spectral weight $Z$
on concentration, and (c) density--density correlation function, $\langle n_{%
\protect\alpha }n_{\protect\alpha ^{\prime }}\rangle $ versus filling, $\bar{%
n}$, in the two--band Hubbard model in $SU(4)$ and for $U=2W=4$. }
\label{FigHUB1}
\end{figure}

If we neglect by the hybridization $\Delta (\omega )$ in Eq. (\ref{H1E}),
the probabilities $X_{n}$ become simply statistical weights: 
\begin{equation}
X_{n}=\frac{e^{-E_{n}/T}}{\sum_{m=0}^{N}C_{m}^{N}e^{-E_{m}/T}} .  \label{STW}
\end{equation}%
We thus see that in principle there are several different ways to determine
the coefficients $X_{n}$, either via self--consistent determination (\ref%
{SCW}), or using statistical formula (\ref{STW}), or taking them from SBMF
equation (\ref{GUZ}), i.e. setting $X_{n}=\psi _{n}^{2}$ while still
utilizing the functional dependence provided by the Hubbard I method. To
determine the best procedure let us first consider limits of large and small 
$U$'s. When $\Delta (\omega )\equiv 0$, $G_{f}(\omega )$ is reduced to $%
\sum_{nm}G_{nm}^{at}(\omega ),$ i.e. the Hubbard\ I method reproduces the
atomic limit. Setting $U\equiv 0$ gives $G_{f}(\omega )=[\omega +\mu
-\epsilon _{f}-\Delta (\omega )]^{-1}$, which is the correct band limit.
Unfortunately, at half--filling this limit has a pathology connected to the
instability towards Mott transition at any interaction strength $U$. To see
this, we consider a dynamical mean--field theory for the Hubbard model.
Using semicircular density of states, we obtain $G_{f}(\omega )=[1-(\frac{W}{%
4})^{2}G_{f}\left( \omega \right) G_{at}(\omega )]^{-1}G_{at}(\omega )$ and
conclude that for any small $U$ the system opens a pathological gap in the
spectrum. Clearly, using Hubbard I only, the behavior of the Green function
at $\omega \rightarrow 0$ cannot be reproduced and the quality of the
numbers $X_{n}$ is at question. This already emphasizes the importance of
using the slave--boson treatment at small frequencies.

Ultimately, making the comparisons with the QMC calculations is the best
option in picking the most accurate procedure to compute the probabilities $%
X_{n}$. To check the accuracy against the QMC we again consider the
two--band Hubbard model in $SU(4)$ symmetry as above. The chemical potential$%
,$ mass renormalization and double occupancy are plotted versus filling in
Fig.~\ref{FigHUB1}. All quantities here were computed with statistical
weights after Eq. (\ref{STW}) but we found a similar agreement while
utilizing the self--consistent determination of $X_{n}$ after Eq. (\ref{SCW}%
). We first see that the Hubbard I approximation does not give satisfactory
agreement with the QMC data for $\bar{n}(\tilde{\mu})$ because it misses the
correct behavior at low frequencies.

The comparisons for $z(\bar{n})$ plotted in Fig. \ref{FigHUB1}(b)
surprisingly show a relatively good behavior. However, the pathology of this
approximation at half--filling would predict $z=0$ for any $U$, which is a
serious warning not to use it for extracting the quasiparticle weight. Fig.~%
\ref{FigHUB1}(c) shows $\langle nn\rangle $ as a function of average
occupation $\bar{n}$. As this quantity is directly related to the
high--frequency expansion one may expect a better accuracy here. However,
comparing Fig. \ref{FigHUB1}(c) and Fig. \ref{FigSBMF}(c), it is clear that
the slave boson method gives more accurate double occupancy. This is due to
the fact that the density matrix obtained by the slave boson method is of
higher quality than the one obtained from the Hubbard I approximation.

The results of these comparisons suggest that the probabilities $\psi
_{n}^{2}$ provided by the slave--boson method is a better way in determining
the coefficients $X_{n}$ in the metallic region of parameters. Therefore it
is preferable to use these numbers while establishing the equations for the
unknown coefficients in the interpolational form (\ref{RAT}). However, the
functional form (\ref{ATG}) of the Hubbard I approximation with $X_{n}=\psi
_{n}^{2}$ can still be used as it provides the positions of the poles $%
P_{n}^{(G)}$ and zeroes $Z_{n}^{(G)}$ of the atomic Green function necessary
for the condition \textit{d)} in the previous Section. This also ensures
accurate high--frequency behavior of the interpolated self--energy since its
moments expressed via multiple occupancies are directly related to $\psi
_{n}^{2}.$

Interestingly, while more sophisticated QMC approach captures both the
quasiparticle peak and the Hubbard bands this is not the case for the
slave--boson mean field method. To obtain the Hubbard bands in this method
fluctuations need to be computed, which would be very tedious in a general
multiorbital situation. However the slave--boson method delivers many
parameters in a good agreement with the QMC results, and, hence, it can be
used to give a functional dependence of the coefficients of the rational
approximation.

\section{\textbf{Results of the Interpolative Scheme}}

By now the procedure to determine the coefficients is well established. We
use the SBMF method to determine $\bar{n},\Sigma (\infty ),$ $\Sigma (0),z$
as well as poles and zeroes of the atomic Green function provided by the
SBMF probabilities $\psi _{n}^{2}$ and by the bare atomic energy levels $%
E_{n}$ (we omit the lifetime effects for simplicity)$.$ This generates a set
of linear equations for coefficients $a_{m}^{\alpha }$, $b_{m}^{\alpha }$ in
the rational interpolation formula (\ref{RAT}). In the present Section we
show the trends our interpolative algorithm gives for the spectral functions
in various regions of parameters as well as provide detailed comparisons for
some values of doping for both imaginary and real axis spectral functions.
The two--band Hubbard model with semicircular density of states and DMFT\
self--consistency condition after (\ref{SCF}) is utilized in $SU(N=4)$
symmetry in all cases using the bandwidth $W=2$ and temperature $T=1/16$.

\subsection{\textit{Trends}}

\begin{figure}[tbp]
\includegraphics[height=0.5\textwidth]{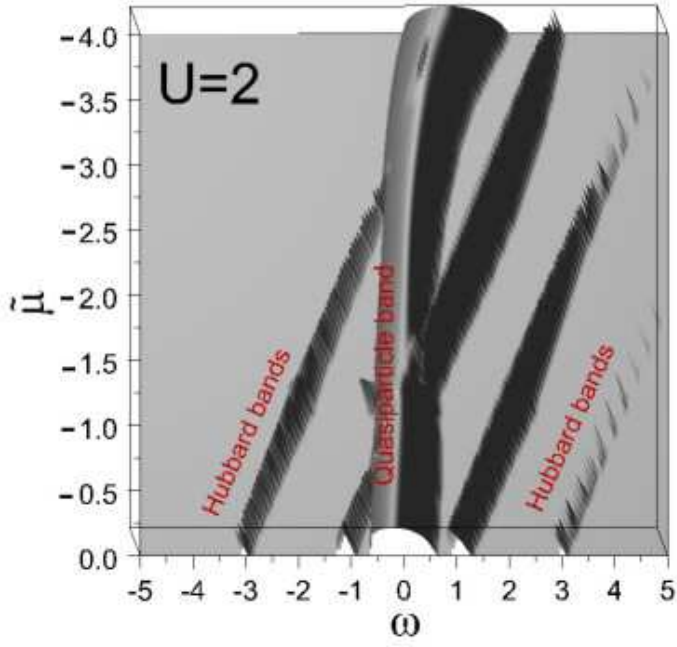}\\[-0.3cm]
\caption{Calculated density of states using the interpolative method as a
function of chemical potential $\tilde{\protect\mu}=\protect\mu -\protect%
\epsilon _{f}-(N-1)U/2$ and frequency for the two--band Hubbard model in $%
SU(4)$ and at $U=W=2$.}
\label{FigU2trend}
\end{figure}

Fig. \ref{FigU2trend} shows the behavior of the density of states $N(\omega
)=-\Im G_{f}(\omega )/\pi $ for $U=W$ as a function of the chemical
potential $\tilde{\mu}$ computed with respect to the particle--hole symmetry
point $(N-1)U/2$ and as a function of frequency $\omega .$ The semicircular
quasiparticle band is seen at the central part of the figure. Its bandwidth
is only weakly renormalized by the interactions in this regime. It is
half--filled for $\tilde{\mu}=0$ (i.e. when $\mu =\epsilon _{f}-(N-1)U/2$)
and gets fully emptied when chemical potential is shifted to negative
values. Several weak satellites can be also seen on this figure which are
due to atomic poles. Their spectral weight is extremely small in this case
and any sizable lifetime effect (which is not included while plotting this
figure) will smear these satellites out almost completely. While approaching
fully emptied (or fully filled situation) the spectral weight of the Hubbard
bands disappears completely and only unrenormalized quasiparticle band
remains. It is clear that even without shifting the atomic poles to the
complex axis, the numerical procedure of generating the self--energy is
absolutely stable.

\begin{figure}[tbp]
\includegraphics[height=0.5\textwidth]{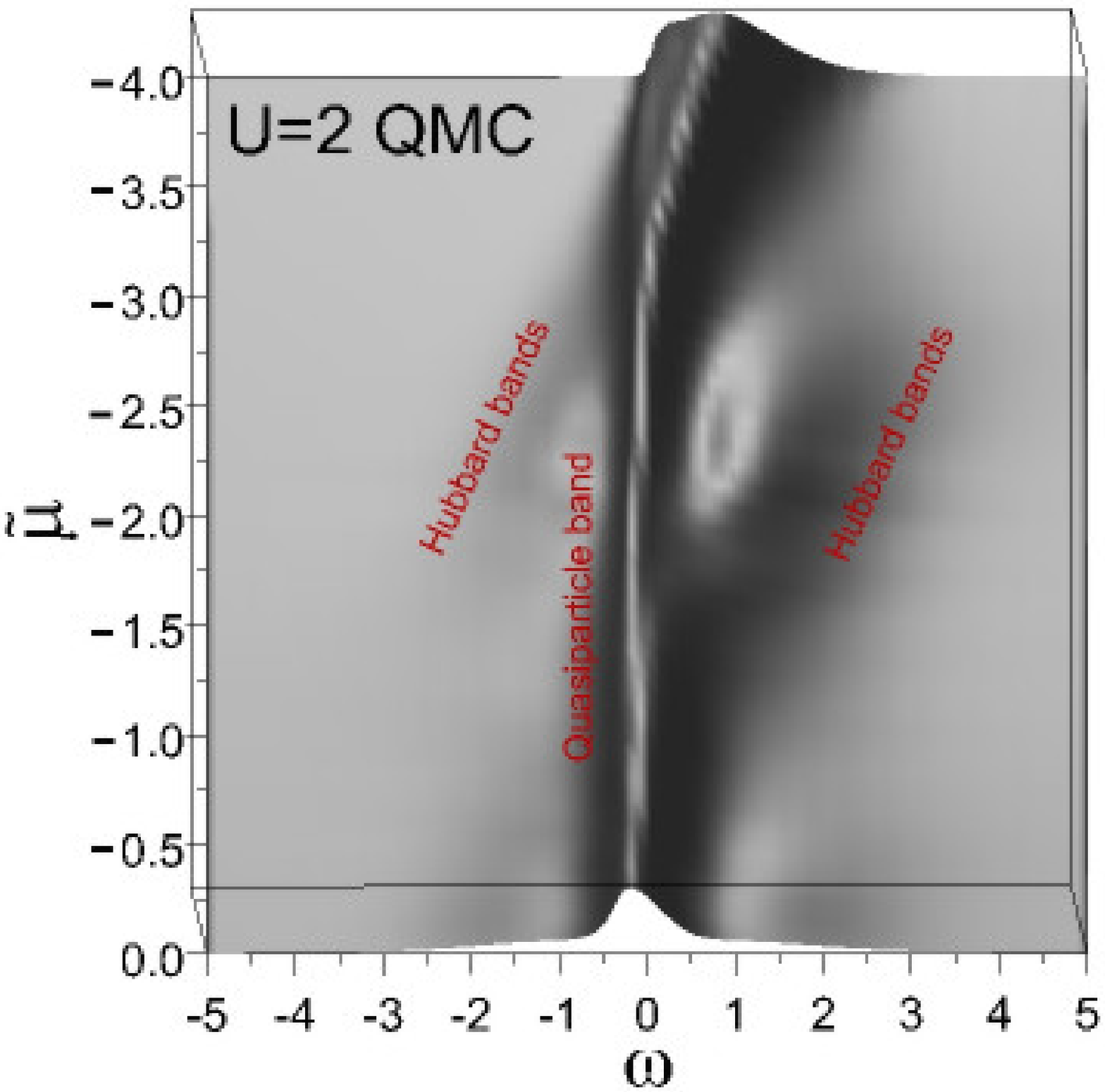}\\[-0.3cm]
\caption{Calculated density of states using the QMC method as a function of
chemical potential $\tilde{\protect\mu}=\protect\mu -\protect\epsilon %
_{f}-(N-1)U/2$ and frequency for the two--band Hubbard model in $SU(4)$\ and
at $U=W=2$. }
\label{FigU2trendQMC}
\end{figure}

\begin{figure}[tbp]
\includegraphics[height=0.5\textwidth]{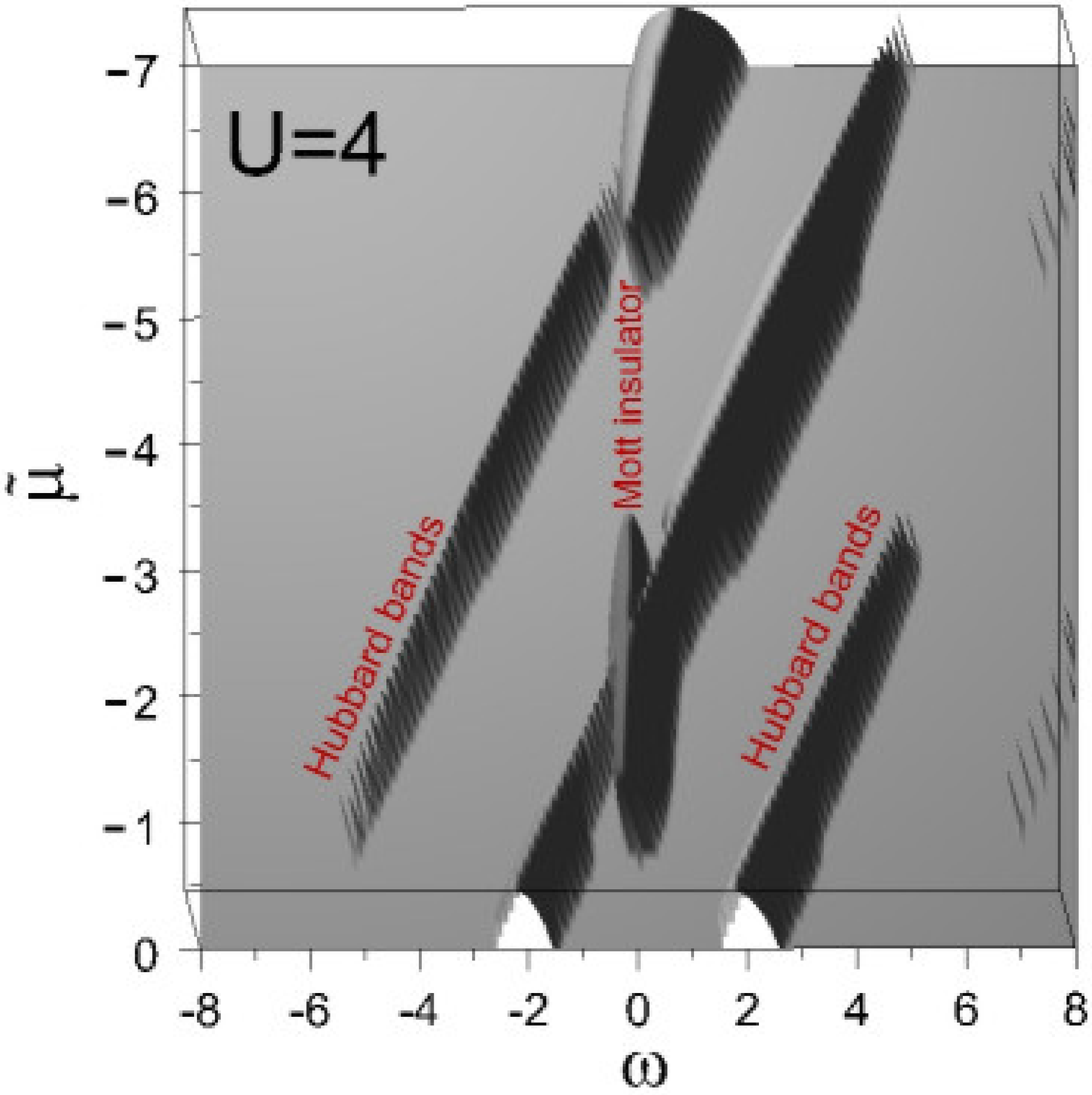}\\[-0.3cm]
\caption{Calculated density of states using the interpolative method as a
function of chemical potential $\tilde{\protect\mu}=\protect\mu -\protect%
\epsilon _{f}-(N-1)U/2$ and frequency for the two--band Hubbard model in $%
SU(4)$ and at $U=2W=4$.}
\label{FigU4trend}
\end{figure}

\begin{figure}[tbp]
\includegraphics[height=0.5\textwidth]{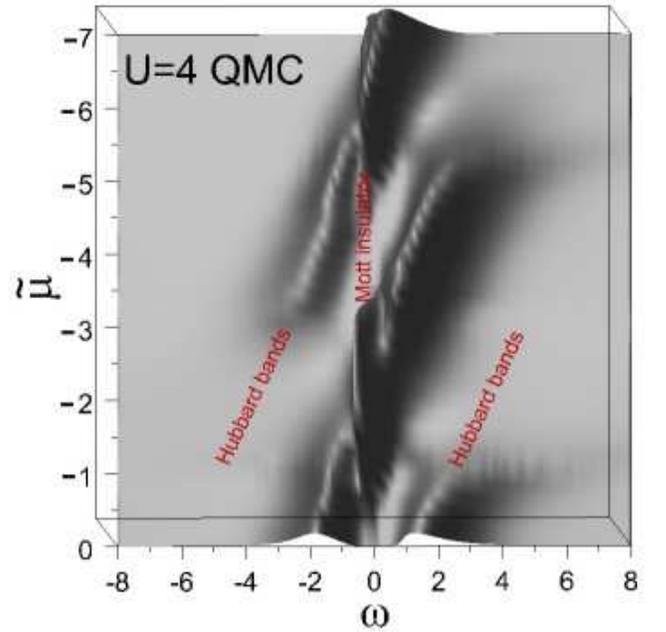}\\[-0.3cm]
\caption{Calculated density of states using the QMC method as a function of
chemical potential $\tilde{\protect\mu}=\protect\mu -\protect\epsilon %
_{f}-(N-1)U/2$ and frequency for the two--band Hubbard model in $SU(4)$\ and
at $U=2W=4$. }
\label{FigU4trendQMC}
\end{figure}

This trend can be directly compared with the simulations using a more
accurate QMC\ impurity solver. We present this in Fig. \ref{FigU2trendQMC}
for $U=W$, which shows calculated density of states in the same region of
parameters. Remarkably that again we can distinguish the renormalized
quasiparticle band and very weak Hubbard satellites. The Hubbard bands
appear to be much more diffuse in this figure mainly due to the lifetime
effects and partially due to maximum entropy method using for analytical
continuation from imaginary to real axis. Otherwise the entire picture looks
very much like the one on Fig. \ref{FigU2trend}, generated with much less
computational effort.

Fig. \ref{FigU4trend} gives the same behavior of the density of states for
the strongly correlated regime $U=2W.$ In this case the situation at integer
filling is totally different as the system undergoes metal--insulator
transition. This is seen around the dopings levels with $\tilde{\mu}$
between 0 and -1 and between -3 and -5 where the wight of the quasiparticle
band collapses while lower and upper Hubbard bands acquire all spectral
weight. In the remaining region of parameters both strongly renormalized
quasiparticle band and Hubbard satellites remain. Again, once full filling
or full emptying is approached the quasiparticle bands restores its original
bandwidth while the Hubbard bands disappear. The QMC\ result for the same
region of parameters is given in Fig. \ref{FigU4trendQMC}. Again we can
distinguish the renormalized quasiparticle band and Hubbard satellites as
well as the areas of Mott insulator and of strongly correlated metal. The
Hubbard bands appear to be more sharp in this figure which signals on
approaching the atomic limit.

\subsection{\textit{Comparison for Spectral Functions at Imaginary Axis }}

We now turn to the comparison of the Green functions and the self--energies
obtained using the formulae (\ref{IMP}), and (\ref{RAT}) respectively
against the predictions of the quantum Monte Carlo method. We will report
our comparisons for the two--band Hubbard model and sets of dopings
corresponding to $\bar{n}=0.5,0.8,1.2,1.5,1.8$ using the value of $U=2W=4.$
Other tests for different degeneracies, doping levels and the interactions
have been performed which display similar accuracy.

\begin{figure}[tbp]
\includegraphics[height=0.65\textwidth]{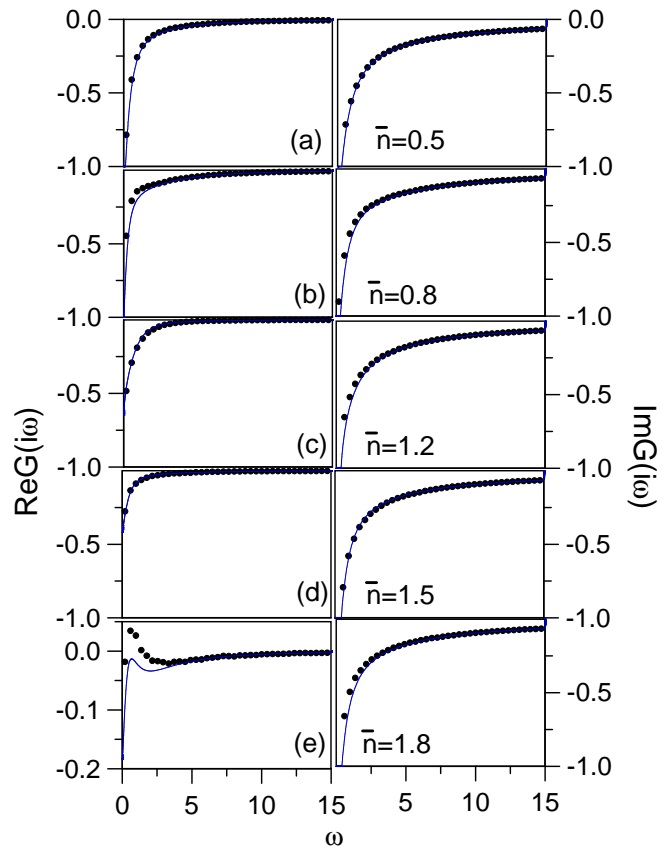}
\caption{Comparison between real and imaginary parts of the Green function
obtained from the interpoaltive method and the quantum Monte Carlo
calculation for the two--band Hubbard model at a set of fillings $\bar{n}%
=0.5,0.8,1.2,1.5,1.8$ and $U=2W=4.$}
\label{FigGRN}
\end{figure}

Fig.~\ref{FigGRN} shows the comparison between the real and imaginary parts
of the Green function obtained by the interpolative method with the results
of the QMC calculations. As one can see almost complete agreement has been
obtained for a wide regime of dopings. The agreement gets less accurate once
the half--filling is approached, but still very good giving an extraordinary
computational speed of the given method compared to QMC.

\begin{figure}[tbp]
\includegraphics[height=0.65\textwidth]{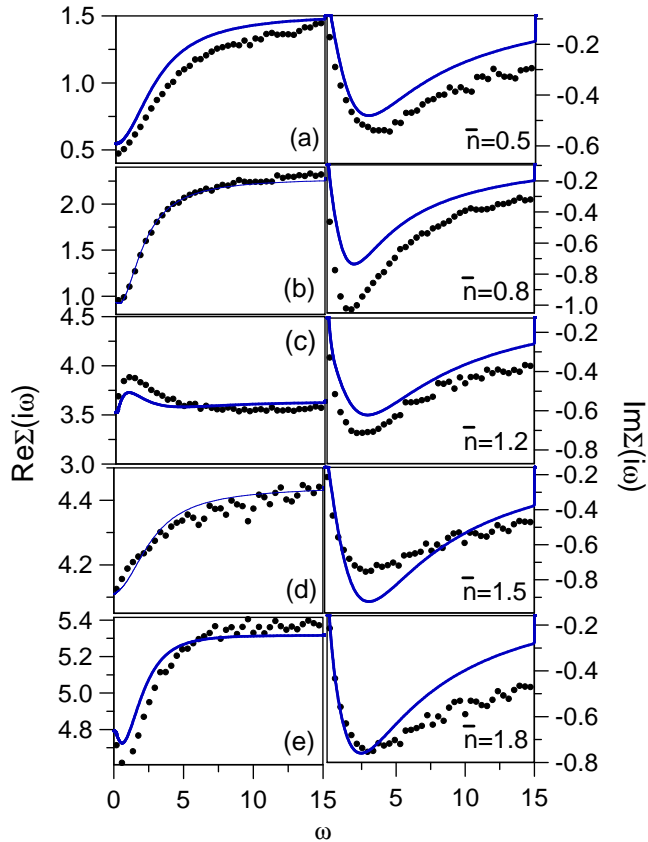}\\[-0.3cm]
\caption{Comparison between real and imaginary parts of the self--energies
obtained from the interpoaltive method and the quantum Monte Carlo
calculation for the two--band Hubbard model at a set of fillings $\bar{n}%
=0.5,0.8,1.2,1.5,1.8$ and $U=2W=4.$}
\label{FigSIG}
\end{figure}

Fig.~\ref{FigSIG} shows similar comparison between the real and imaginary
parts of the self--energies obtained by the interpolative and the QMC
method. We can see that the self--energies exhibit some noise which is
intrinsic to stochastic QMC procedure. The values of the self-energies near $%
i\omega =0$ and $i\omega =\infty $ are correctly captured with some residual
discrepancies are attributed to slightly different chemical potentials used
to reproduce given filling within every method. The results at the imaginary
axis show slightly underestimated slopes of the self--energies within the
interpolative algorithm which is attributed to the underestimated values of $%
z$ obtained from the SBMF calculation. Ultimately improving these numbers by
inclusions of fluctuations beyond mean field will further improve the
comparisons. However, even at the present stage of the accuracy all
functional dependence brought by the SBMF method quantitatively captures the
behavior of the self--energy seen from time consuming QMC simulation.

\subsection{\textit{Comparison for Spectral Functions at Real Axis }}

\begin{figure}[tbp]
\includegraphics[height=0.65\textwidth]{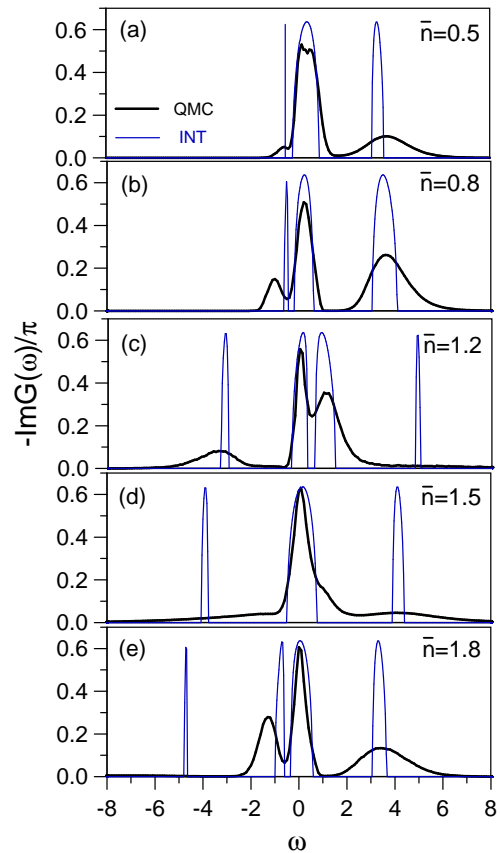}\\[-0.3cm]
\caption{Comparison between\ the one--electron densities of states obtained
from interpolative formula and the quantum Monte Carlo calculation for the
two--band Hubbard model at fillings $\bar{n}=0.5,0.8,1.2,1.5,1.8$. and $%
U=2W=4.$}
\label{FigDOS}
\end{figure}

We also made detailed comparisons between calculated densities of states
obtained at the real axis using the interpolative method and the QMC\
algorithm. The QMC\ densities of states require an analytical continuation
from the imaginary to real axis and were generated using the maximum entropy
method. By itself this procedure introduces some errors within the QMC
especially at higher frequencies. In Fig.~\ref{FigDOS}, we show our results
for the fillings corresponding to $\bar{n}=0.5,0.8,1.2,1.5,1.8$ using the
value of $U=2W=4$. One can see the appearance of the quasiparticle band and
two Hubbard bands distanced by the value of $U$. It can be seen that the
interpolative method remarkably reproduces the trend in shifting the Hubbard
bands upon changing the doping. It automatically holds the distance between
them to the value of $U$ while this is not always true in the quantum Monte
Carlo method. Despite this result, the overall agreement between both
methods is very satisfactory.

\section{Discussion}

Here we would like to discuss possible ways to further improve the accuracy
of the method. The inaccuracies are mainly seen in three different
quantities: i) the width of the Hubbard bands, ii) the mass renormalization $%
z(\mu )$ which is borrowed from the SBMF method, and iii) the number of
electrons $n(\mu )$ extracted from the interpolated impurity Green function (%
\ref{IMP}). The inaccuracy in the width of the Hubbard band is mainly
connected to neglecting the lifetime effect. Provided it is computed, this
will shift the positions of atomic poles onto the complex plane which is in
principle trivial to account for within our interpolative algorithm. To
improve the accuracy of $z(\mu )$ one can, for example, work out a modified
slave boson scheme which will account for the fluctuations around mean field
solution. The inaccuracy in $n(\mu )$ is small in many regions of parameters
and typically amounts to 5--10 per cent. We can try to improve this
agreement by requirement that $n(\mu )$ obtained via interpolation matches
with $n_{SBMF}(\mu )$ obtained by the SBMF method. The latter agrees very
well with the QMC for a wide region of parameters as it is evident from Fig. %
\ref{FigSBMF}(a). In reality, our analysis shows that in many cases the
discrepancy in $n(\mu )$ is connected with the overestimation of $%
z_{SBMF}(\mu )$. Therefore, points ii) and iii) mentioned above are
interrelated.

The requirement that $n(\mu )=n_{SBMF}(\mu )$ can be enforced by adjusting
the width of the quasiparticle band, and in many regions of parameters this
is controlled by $z(\mu ).$ However, there are situations when the Hubbard
band appears in the vicinity of $\omega =0$, and changing $z(\mu )$ does not
affect the bandwidth$.$ To gain a control in those cases it is better to
replace the constraints $\Sigma (0),d\Sigma /d\omega |_{\omega =0}$ by
constraints of fixing the self--energies at two frequencies$,$ say $\Sigma
(0)$ and $\Sigma (i\omega _{0})$ where $\omega _{0}$ is the frequency of the
order of renormalized bandwidth$.$ We have found that this scheme brings
mass renormalizations which are about 30\% smaller than the SBMF ones, and
the agreement with the QMC is significantly improved. Thus, inaccuracies ii)
and iii) can be avoided with this very cheap trick. However, we also would
like to point out that the condition $n(\mu )=n_{SBMF}(\mu )$ is essentially
non--linear as the solution may not exist for all regions of parameters. It
is, for example, evident that in such points where $n(\mu )$ is given by a
symmetry (as, e.g., particle--hole symmetry point $n=2$ in the case
considered above) the mass renormalization does not affect the number of
electrons.

As the philosophy of our approach is to get the best possible fit we are
also open to implementing any kinds of \textit{ad hoc} renormalizations
constants. One of such possibility could be the use of a quasiparticle
residue 30 per cent smaller than $z_{SBMF}(\mu ).$ As $z(\mu )$ should go to
unity when $U=0$, the correction can, for example, be encoded into the
formula $z(\mu )=z_{SBMF}(\mu )[0.7+0.3z_{SBMF}(\mu )].$

We finally would like to remark that the scheme defined by a set of linear
equations for the coefficients (\ref{CO1})--(\ref{CO6}) is absolutely robust
as solutions exist for all regimes of parameters such as strength of the
interaction, doping and degeneracy. In general, bringing any information on
the self energy $\Sigma (\omega _{x})$ at some frequency point $\omega _{x}$
or its derivative $d\Sigma /d\omega |_{\omega =\omega _{x}}$ would generate
a linear relationship between the interpolation coefficients, thus keeping
robustness of the method. On the other hand, fixing such relationships as
numbers of electrons brings non--linearity to the problem which could lead
to multiplicity or non--existence of the solutions. It is also clear that by
narrowing the regime of parameters, the accuracy of the interpolative
algorithm can be systematically increased.

\section{Conclusion}

To summarize, this paper shows the possibility to interpolate the
self--energies for a whole range of dopings, degeneracies and the
interactions using a computationally efficient algorithm. The parameters of
the interpolation are obtained from a set of constraints in the slave boson
mean field method combined with the functional form of the atomic Green
function. The interpolative method reproduces all trends in remarkable
agreement with such sophisticated and numerically accurate impurity solver
as the QMC method. We also obtain a very good quantitative agreement in a
whole range of parameters for such quantities as mean level occupancies,
spectral functions and self--energies. Some residual discrepancies remain
which can be corrected provided better algorithms delivering the constraints
will be utilized. Nevertheless, given the superior speed of the present
approach, we have obtained a truly exceptional accuracy times efficiency of
the proposed procedure.

The work was supported by NSF--DMR Grants 0096462, 02382188, 0312478,
0342290 and US DOE Grant No DE--FG02--99ER45761. The authors also
acknowledge the financial support from the Computational Material Science
Network operated by US DOE and from the Ministry of Education, Science and
Sport of Slovenia.

\begin{widetext}

\section{Appendix}

In the crystal field case we assume that $N$--fold degenerate impurity level
$\epsilon _{f}$ is split by a crystal field onto $G$ sublevels $%
\epsilon_{f1},...\epsilon_{f\alpha },...\epsilon_{fG}$. We assume that for
each sublevel there is still some partial degeneracy $d_{\alpha }$ so that $%
\sum_{\alpha =1}^{G}d_{\alpha }=N.$ In limiting case of $SU(N)$ degeneracy, $%
G=1,d_{1}=N$, and in non--degenerate case, $G=N,d_{1}=d_{\alpha }=d_{G}=1$.
We need to discuss how a number of electrons $n$ can be accommodated over
different sublevels $\epsilon_{f\alpha }$. Introducing numbers of electrons
on each sublevel, $n_{\alpha }$, we obtain $\sum_{\alpha =1}^{G}n_{\alpha
}=n.$ Note the restrictions: $0<n<$ $N,$ and $0<n_{\alpha }\leq d_{\alpha }$%
. In $SU(N)$ case, $G=1,n_{1}=n$, and in non--degenerate case, $G=N$, $%
n_{\alpha }$ is either 0 or 1. Total energy for the shell with $n$ electrons
depends on particular configuration $\{n_{\alpha }\}$
\begin{equation}
E_{n_{1}...n_{G}}=\sum_{\alpha =1}^{G}\epsilon _{f\alpha }n_{\alpha }+\frac{%
1 }{2}U(\Sigma _{\alpha }n_{\alpha })[(\Sigma _{\alpha }n_{\alpha })-1].
\label{SLAcrfLEV}
\end{equation}
Many--body wave function is also characterized by a set of numbers $%
\{n_{\alpha }\}$, i.e. $|n_{1}...n_{G}\rangle .$ Energy $E_{n_{1}...n_{G}}$
remains degenerate, which can be calculated as product of how many
combinations exists to accommodate electrons in each sublevel, i.e. $%
C_{n_{1}}^{d_{1}}\times ...C_{n_{\alpha }}^{d_{\alpha }}\times
...C_{n_{G}}^{d_{\alpha }}.$ Let us further introduce probabilities $\psi
_{n_{1}...n_{G}}$ to find a shell in a given state with energy $%
E_{n_{1}...n_{G}}.$ Sum of all probabilities should be equal to $1$, i.e.%
\begin{equation}
\sum_{n_{1}=0}^{d_{1}}...\sum_{n_{\alpha }=0}^{d_{\alpha
}}...\sum_{n_{G}=0}^{d_{G}}C_{n_{1}}^{d_{1}}...C_{n_{\alpha }}^{d_{\alpha
}}...C_{n_{G}}^{d_{G}}\psi _{n_{1}...n_{G}}^{2}=1.  \label{SLAcrfSUM}
\end{equation}

There are two Green functions in Gutzwiller method: impurity Green function $%
\hat{G}_{f}(i\omega )$ and quasiparticle Green function $\hat{G}_{g}(\omega
)=\hat{b}^{-1}\hat{G}_{f}(\omega )\hat{b}^{-1},$ where matrix coefficients $%
\hat{b}$ represent generalized mass renormalizations parameters. All
matrices are assumed to be diagonal and have diagonal elements enumerated as
follows: $G_{1}(\omega ),...G_{\alpha }(\omega ),...G_{G}(\omega ).$ Each
element in the Green function is represented as follows%
\begin{equation}
G_{g\alpha }(\omega )=\frac{1}{\omega -\lambda _{\alpha }-b_{\alpha
}^{2}\Delta _{\alpha }(\omega )},  \label{SLAcrfGRF}
\end{equation}%
\begin{equation}
G_{f\alpha }(\omega )=b_{\alpha }^{2}G_{g\alpha }(\omega ).
\label{SLAcrfGRG}
\end{equation}%
and determines a mean number of electrons in each sublevel%
\begin{equation}
\bar{n}_{\alpha }=d_{\alpha }T\sum_{i\omega }G_{g\alpha }(i\omega
)e^{i\omega 0^{+}}.  \label{SLAcrfNAV}
\end{equation}%
The total mean number of electrons is thus: $\bar{n}=\sum_{\alpha =1}^{G}%
\bar{n}_{\alpha }.$ Hybridization function $\hat{\Delta}(i\omega )$ is the
matrix which is assumed to be diagonal, and it has diagonal elements enumerated as
follows: $\Delta _{1}(\omega ),...\Delta _{\alpha }(\omega ),...\Delta
_{G}(\omega ).$ Mass renormalizations $Z_{\alpha }=b_{\alpha }^{2}$ are
determined in each sublevel.

Diagonal elements for the self--energy are
\begin{equation}
\Sigma _{\alpha }(\omega )=\omega +\mu -\epsilon _{f\alpha }-\Delta _{\alpha
}(\omega )-G_{\alpha }^{-1}(\omega )=\omega (1-\frac{1}{b_{\alpha }^{2}}%
)-\epsilon _{f\alpha }-\frac{\lambda _{\alpha }}{b_{\alpha }^{2}}.
\label{SLAcrfSIG}
\end{equation}%
Here:%
\begin{eqnarray}
b_{\alpha } &=&R_{\alpha }L_{\alpha
}\sum_{n_{1}=0}^{d_{1}}...\sum_{n_{\alpha }=1}^{d_{\alpha
}}...\sum_{n_{G}=0}^{d_{G}}C_{n_{1}}^{d_{1}}...C_{n_{\alpha }-1}^{d_{\alpha
}-1}...C_{n_{G}}^{d_{G}}\psi _{n_{1}...n_{\alpha }...n_{G}}\psi
_{n_{1}...n_{\alpha }-1...n_{G}},  \label{SLAcrfEQB} \\
L_{\alpha } &=&\left( 1-\sum_{n_{1}=0}^{d_{1}}...\sum_{n_{\alpha
}=1}^{d_{\alpha }}...\sum_{n_{G}=0}^{d_{G}}C_{n_{1}}^{d_{1}}...C_{n_{\alpha
}-1}^{d_{\alpha }-1}...C_{n_{G}}^{d_{G}}\psi _{n_{1}...n_{\alpha
}...n_{G}}^{2}\right) ^{-1/2},  \label{SLAcrfEQL} \\
R_{\alpha } &=&\left( 1-\sum_{n_{1}=0}^{d_{1}}...\sum_{n_{\alpha
}=0}^{d_{\alpha
}-1}...\sum_{n_{G}=0}^{d_{G}}C_{n_{1}}^{d_{1}}...C_{n_{\alpha }}^{d_{\alpha
}-1}...C_{n_{G}}^{d_{G}}\psi _{n_{1}...n_{\alpha }...n_{G}}^{2}\right)
^{-1/2}.  \label{LSAcrfEQR}
\end{eqnarray}

The generalization of the non--linear equations (\ref{GUZ}) has the form%
\begin{eqnarray}
0 &=&\left[ E_{n_{1}...n_{G}}+\Lambda -(\Sigma _{\alpha }^{G}\lambda
_{\alpha }n_{\alpha })\right] \psi _{n_{1}...n_{G}}+  \notag \\
&&\sum_{\alpha =1}^{G}n_{\alpha }[T\Sigma _{i\omega }\Delta _{\alpha
}(i\omega )G_{g\alpha }(i\omega )]b_{\alpha }\left[ R_{\alpha }L_{\alpha
}\psi _{n_{1}...n_{\alpha }-1...n_{G}}+b_{\alpha }L_{\alpha }^{2}\psi
_{n_{1}...n_{\alpha }...n_{G}}\right] +  \notag \\
&&\sum_{\alpha =1}^{G}(d_{\alpha }-n_{\alpha })[T\Sigma _{i\omega }\Delta
_{\alpha }(i\omega )G_{g\alpha }(i\omega )]b_{\alpha }\left[ R_{\alpha
}L_{\alpha }\psi _{n_{1}...n_{\alpha }+1...n_{G}}+b_{\alpha }R_{\alpha
}^{2}\psi _{n_{1}...n_{\alpha }...n_{G}}\right] .  \label{SLAcrfSCF}
\end{eqnarray}

\end{widetext}


\begin{thebibliography}{99}
\bibitem{DMFT} For a review, see, A. Georges, G. Kotliar, W. Krauth, and
M.~Rozenberg, Rev. Mod. Phys. \textbf{68}, 13 (1996).

\bibitem{LDA} For a review, see, \textit{e.g.}, \textit{Theory of the
Inhomogeneous Electron Gas}, edited by S. Lundqvist and S. H. March (Plenum,
New York, 1983).

\bibitem{AnisimovKotliar} V. I. Anisimov, A. I. Poteryaev, M. A. Korotin, A.
O. Anokhin, and G. Kotliar, J.\ Phys.: Condens.\ Matter \textbf{35}, 7359
(1997).

\bibitem{LDA++} A. Lichtenstein and M. Katsnelson, Phys. Rev. B \textbf{57},
6884 (1998).

\bibitem{Georges} G. Kotliar and S. Y. Savrasov, in \textit{New Theoretical
approaches to strongly correlated systems}, edited by A. M. Tsvelik, (Kluwer
Academic Publishers, the Netherlands, 2001), p. 259, (available in
cond--mat/020824); S. Biermann, F. Aryasetiawan, and A. Georges, Phys. Rev.
Lett. \textbf{90}, 086402 (2003).

\bibitem{SDFT} S. Y. Savrasov and G. Kotliar, Phys. Rev. B \textbf{69},
245101 (2004).

\bibitem{LaTiO3} I.A. Nekrasov, K. Held, N. Blumer, A.I. Poteryaev, V.I.
Anisimov, and D. Vollhardt, Eur.\ Phys.\ J.\ B\textbf{18}, 55 (2000).

\bibitem{V2O3} K. Held, G. Keller, V. Eyert, D. Vollhardt, and V.I.
Anisimov, Phys. Rev. Lett. \textbf{86}, 5345 (2001).

\bibitem{Licht} A. I. Lichtenstein, M. I. Katsnelson, and G. Kotliar, Phys.
Rev. Lett. \textbf{87}, 067205 (2001).

\bibitem{Ce} K. Held, A.K. McMahan, and R.T. Scalettar, Phys. Rev. Lett. 
\textbf{87}, 276404 (2001).

\bibitem{Nature} S. Savrasov, G. Kotliar, and E. Abrahams, Nature \textbf{410%
}, 793 (2001).

\bibitem{Science} Xi Dai, S. Savrasov, G. Kotliar, A. Migliori, H. Ledbetter
and E. Abrahams, Science \textbf{300}, 953 (2003).

\bibitem{NiOPRL} S. Y. Savrasov and G. Kotliar, Phys. Rev. Lett \textbf{90},
056401 (2003).

\bibitem{reviews} K. Held, I. A. Nekrasov, G. Keller, V. Eyert, N. Bluemer,
A. K. McMahan, R. T. Scalettar, Th. Pruschke, V. I. Anisimov, and D.
Vollhardt, Psi--k Newsletter \#\textbf{56} (April 2003), p. 65; A. I.
Lichtenstein, M. I. Katsnelson, and G. Kotliar, in \textit{Electron
Correlations and Materials Properties}, ed. by A. Gonis, N. Kioussis and M.
Ciftan (Kluwer Academic, Plenum Publishers, 2002) p. 428.

\bibitem{AIM} P. W. Anderson, Phys. Rev. \textbf{124}, 41 (1961).

\bibitem{Jarrell} For a review, see, e.g., M. Jarrell, and J.\ E.\
Gubernatis, Physics Reports, \textbf{269}, 133 (1996).

\bibitem{jeschke} H. Jeschke and G. Kotliar, Rutgers University preprint,
2003.

\bibitem{rotors} S. Florens and A. Georges Phys. Rev. B \textbf{66}, 165111
(2002).

\bibitem{Haule:2001} K. Haule, S. Kirchner, J. Kroha, and P. W\"{o}lfle,
Phys. Rev. B \textbf{64}, 155111 (2001).

\bibitem{Vosko} S. H. Vosko, L. Wilk, and M. Nusair, Can. J. Phys. \textbf{58%
}, 1200 (1980).

\bibitem{Ceperley} D. M. Ceperley and B. J. Alder, Phys. Rev. Lett. \textbf{%
45}, 566 (1980).

\bibitem{Gutz} M. Gutzwiller, Phys. Rev. \textbf{134}, A923 (1964).

\bibitem{Ruck} G. Kotliar and A. E. Ruckenstein, Phys. Rev. Lett. \textbf{57}%
, 1362 (1986).

\bibitem{Fleszar} R. Fresard and G. Kotliar, Phys. Rev. B. \textbf{56},
12909 (1997).

\bibitem{Hasegawa} H. Hasegawa, Phys. Rev. B \textbf{56}, 1196 (1997).

\bibitem{Hubbard1} J. Hubbard, Proc. Roy. Soc. (London) A\textbf{281}, 401
(1964).

\bibitem{Bulla} R. Bulla, T. A. Costi, and D. Vollhardt, Phys.\ Rev. B 
\textbf{64}, 045103 (2001).

\bibitem{Florens} S. Florens, A. Georges, G. Kotliar, and O. Parcollet,
Phys. Rev. B \textbf{66}, 205102 (2002).

\bibitem{Broyden} See, e.g., D. D. Johnson, Phys. Rev. B \textbf{38}, 12807
(1988).

\bibitem{Moments} W. Nolting, and W. Borgiel, Phys. Rev. B \textbf{39},
6962(1989).

\bibitem{Roth} L. M. Roth, Phys. Rev. \textbf{184}, 451 (1969).

\bibitem{Pade} H. J.\ Vidberg and J. W.\ Serene, Journal of Low Temperature
Physics, \textbf{29}, 179 (1977).
\end{thebibliography}
\end{document}